\documentstyle[11pt,newpasp,twoside,psfig]{article}
\index{summary}
\index{instructions}
\index{template}

\def\lapp{\ifmmode\stackrel{<}{_{\sim}}\else$\stackrel{<}{_{\sim}}$\fi}
\def\gapp{\ifmmode\stackrel{>}{_{\sim}}\else$\stackrel{>}{_{\sim}}$\fi}

\def\edcomment#1{\iffalse\marginpar{\raggedright\sl#1\/}\else\relax\fi}
\reversemarginpar

\begin{document}
\title{X-ray Detection of PSR B1757$-$24 and its Nebular Tail}
\author{V. M. Kaspi}
\affil{McGill University, Rutherford Physics Building, 3600 University Street, Montreal, QC Canada H3A 2T8}
\affil{Department of Physics and Center for Space Research, Massachusetts Institute of Technology, 70 Vassar Street, Cambridge, MA 02139}
\author{E. V. Gotthelf}
\affil{Columbia University Astronomy Department, Pupin Hall, 550 West 120th Street, New York, NY 10027}
\author{B. M. Gaensler}
\affil{Harvard-Smithsonian Center for Astrophysics, 60 Garden Street, Cambridge, MA 02138}
\author{M. Lyutikov}
\affil{McGill University, Rutherford Physics Building, 3600 University Street, Montreal,
 QC Canada H3A 2T8}
\affil{Department of Physics and Center for Space Research, Massachusetts Institute of Technology, 70 Vassar Street, Cambridge, MA 02139}

\begin{abstract}
We report the first X-ray detection of the radio pulsar PSR B1757$-$24
using the {\it Chandra X-ray Observatory}.  The image reveals
point-source emission at the pulsar position, consistent with being
magnetospheric emission from the pulsar.  In addition, we detect a
faint tail extending nearly 20$''$ east of the pulsar, in the same
direction and with comparable morphology to the pulsar's 
well-studied radio tail.
The X-ray tail is unlikely to be emission left behind following the
passage of the pulsar, but rather is probably from synchrotron-emitting pulsar
wind particles having flow velocity $\sim$7000~km~s$^{-1}$.  Assuming
the point-source X-ray emission is magnetospheric, the observed X-ray
tail represents only $\sim$0.01\% of the pulsar's spin-down
luminosity, significantly lower than the analogous
efficiencies of most known X-ray nebulae surrounding rotation-powered
pulsars.
\end{abstract}

\section{Introduction}

PSR~B1757$-$24 is a 124-ms radio pulsar near the supernova remnant
(SNR) G5.4$-$1.2.  Radio timing observations show that the pulsar has
characteristic age 16~kyr and spin-down luminosity $\dot{E} = 2.6
\times 10^{36}$~erg~s$^{-1}$ (Manchester et al. 1991).  The pulsar is
at the tip of a flat-spectrum radio protuberance just outside the west
side of the SNR shell.  The protuberance consists of a small, roughly
spherical nebula, G5.27$-$0.90, that has a highly collimated finger
of emission on its
western side.  The pulsar is at the westernmost tip of the finger, whose
morphology and spectrum suggest a ram-pressure confined
pulsar wind nebula (PWN) (Frail \& Kulkarni 1991).  Assuming the pulsar
was born at the center of G5.4$-$1.2 and that the characteristic pulsar
age is a good estimate for the age of a system, the pulsar appears to
have overtaken the expanding shell, implying a transverse space
velocity of $v_t \sim$1800~km~s$^{-1}$ for a distance of 5~kpc (Frail,
Kassim, \& Weiler 1994).  However, recent interferometric observations
failed to detect the implied proper motion (Gaensler \& Frail 2000).
They set a 5$\sigma$ upper limit of $v_t <
590$~km~s$^{-1}$.  This suggests that the pulsar is older than its
characteristic age, or that the assumed pulsar birth place is
incorrect.  

A ram-pressure confined wind should
radiate X-rays as part of the broad-band synchrotron spectrum that
results from the shock-acceleration and subsequent gyration of
relativistic wind electron/positron pairs in the ambient magnetic
field.  This is in contrast to static PWNe in which a high external
pressure, usually supplied by the hot gas interior of a SNR, does the
confining.  The PSR~B1757$-$24 PWN is of interest as it exhibits the
cleanest and most extreme bow-shock plus tail morphology of any such
system, suggesting that it may exemplify the class of ram-pressure
confined PWN most ideally.  We summarize here the first X-ray detection
of PSR~B1757$-$24.  Details of this result can be found in Kaspi et al. (2001).

\section{Observations}

The PSR~B1757$-$24 field was observed on 2000 April 12 for 19.6~ks using
the S3 back-illuminated chip on the Advanced CCD Imaging Spectrometer
(ACIS) instrument aboard the {\it Chandra X-ray Observatory}.  The ACIS
image (Fig. 1, left) reveals a faint point source near the radio pulsar
position, and a fainter tail of emission on the eastern side, similar
to the radio emission.  Figure 1 (right) shows the distribution of
counts as a function of distance from the point source.  With respect
to the mean and the western side, the eastern side clearly has an
overdensity of counts extending nearly 20$''$ from the point source.
Careful astrometry confirms the coincidence between the X-ray point
source and the radio pulsar.

The spectrum of the point source is well characterized  with both an
absorbed power-law model, as well as an absorbed thermal bremsstrahlung
model.  The spectrum of the fainter tail emission is not well
determined, however it is suggested of relatively hard emission, having
power-law index of $\sim$1.  

\begin{figure}[h]
\centerline{
\psfig{file=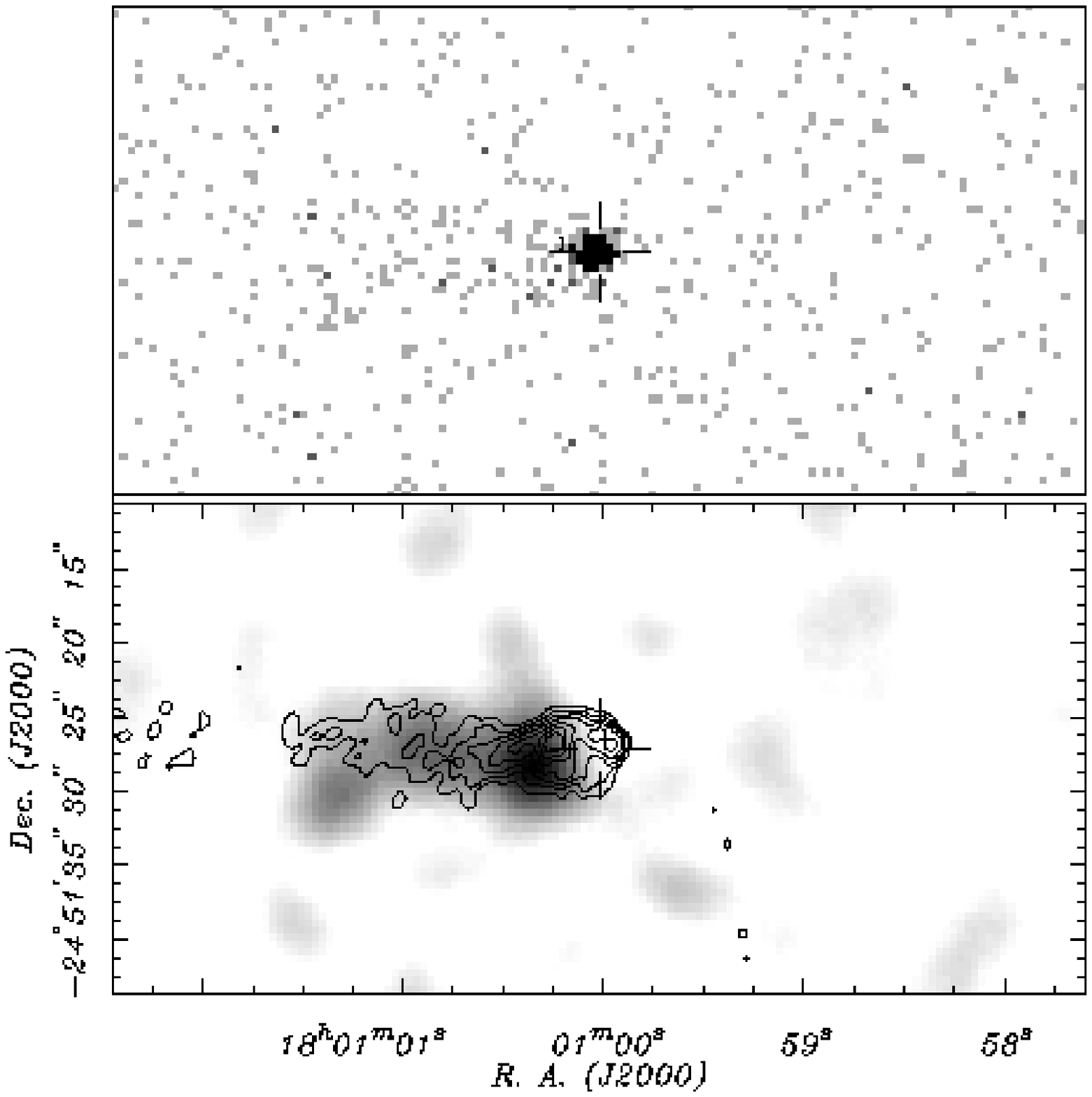,height=3in}
\psfig{file=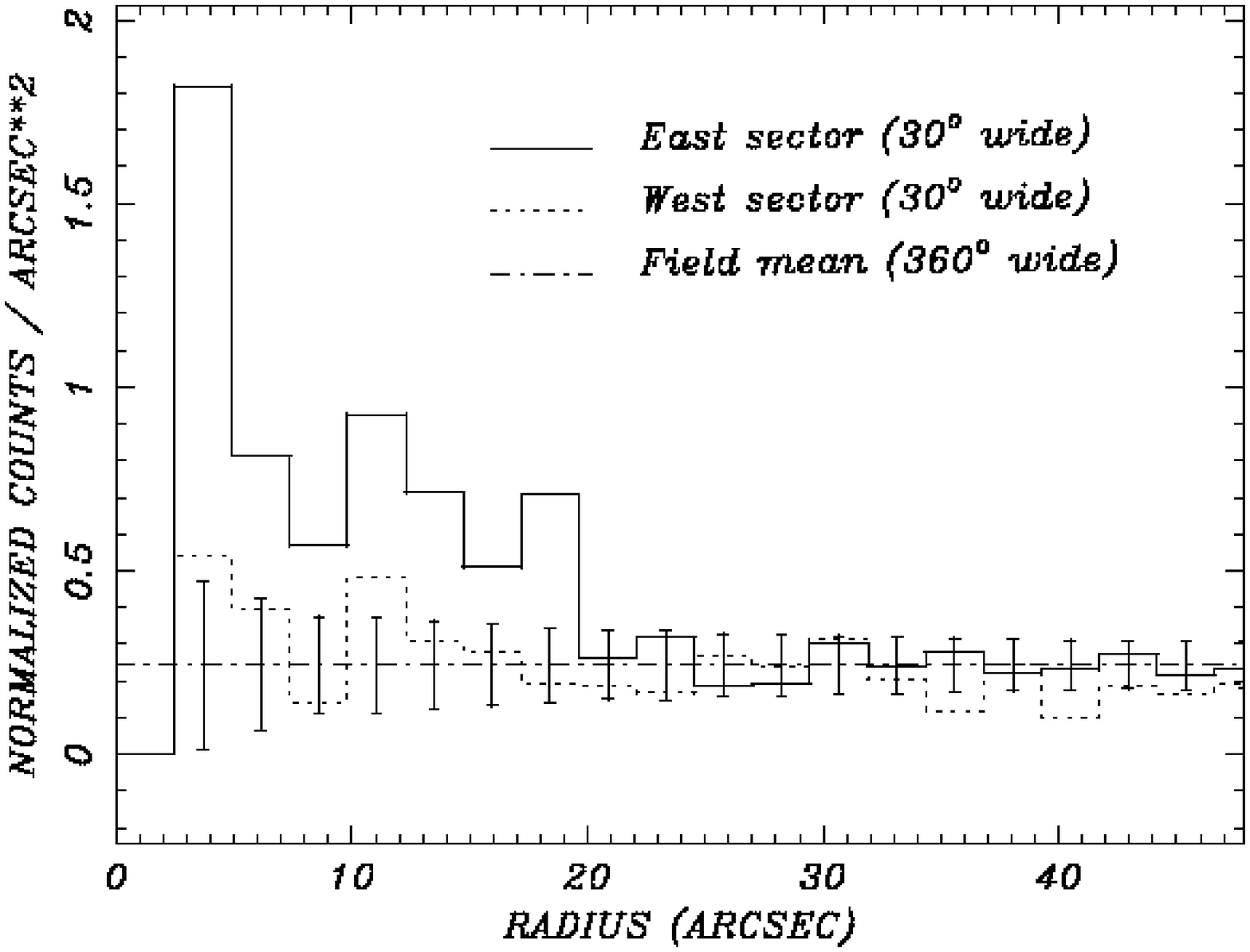,height=3in}}
\caption{Left: (Top) Image in $0.3-10$~keV range.  The
X-ray point source is at the pulsar position (the cross).
(Bottom) Same image
with point source contribution removed, $\sim 2''$ Gaussian
smoothing, and intensity scaled to emphasize diffuse X-ray emission.
Contours show 4.9~GHz radio emission in equally spaced contours ranging
from 0.17 to 0.56~mJy/beam (beam size 1$''$.2). 
Right:  X-ray intensity vs. radius from the point source in two
$30^{\circ}$ wedges, east (solid line) and west (dotted line) of the
pulsar.  The point-source emission has been subtracted.  The field mean
is shown by a dot-dashed line.} 
\end{figure}

\section{Results and Discussion}

The X-ray point source is likely to be non-thermal pulse-phase-averaged
magnetospheric emission from the radio pulsar itself.  The observed
2--10~keV unabsorbed luminosity, for a
distance $d=5$~kpc is $2 \times 10^{33}$~erg~s$^{-1}$, assuming
beaming angle $\phi = \pi$~sr (see Kaspi et al. 2001).  This implies an efficiency of
conversion of spin-down luminosity into magnetospheric emission of
$0.00020(\phi/\pi \; {\rm sr})(d/5 \; {\rm kpc})^2$.  This efficiency,
as well as the measured power-law photon index, are consistent with
those observed for the magnetospheric components of other radio pulsars
(Becker \& Tr\"umper 1997).

The tail X-ray emission is likely to be synchrotron radiation from the
shocked pulsar wind.  The observed X-ray tail extends nearly 20$''$, or
0.48~pc for a distance of 5~kpc, to the east of the pulsar, nearly as
long as the detected radio tail.  The upper limit on the transverse
velocity of $v_t <590$~km~s$^{-1}$ implies the time since the pulsar
was at the eastern-most tip of the observed X-ray emission must be $>
800$~yr.  The synchrotron lifetime of a photon of energy $E$ (in keV)
in a magnetic field $B_{-4}$ (in units of $10^{-4}$~G) is $t_s \simeq
40 E^{-1/2} B_{-4}^{-3/2}$~${\rm yr}$.  Thus, for $t_s > 800$~yr and
$E\simeq 1-9$~keV, $B< 0.8-14 \; \mu$G.  This is much less than the
equipartition magnetic field  $B_{\rm eq} \sim \sqrt{\dot{E}/r_s^2 c }
\sim 70 \; \mu$G expected in the vicinity of the pulsar.  Here, $r_s$
is the distance from the pulsar to the bow shock head.  Hence, the
X-ray tail behind PSR~B1757$-$24 cannot be synchrotron emission from
pulsar wind particles just  left behind after the passage of the
pulsar.  Rather, freshly shocked wind particles must be continuously
fed eastward with a velocity much larger than the pulsar space
velocity, $v_f \gg v_t$.  We can constrain the flow velocity $v_f$ of
the wind particles in the tail by noting that it must be high enough to
continuously supply particles given their cooling times.  Thus, the
flow time $t_f \lapp t_s$.  Assuming  that the magnetic field reaches
its equipartition value near the bow-shock head and that this value
holds for the approximately
 one-dimensional  tail region too, we find $t_s \gapp 70 (E / 1 \; {\rm
keV})^{-1/2} (B_{\rm eq} / 70 \; \mu{\rm G})^{-3/ 2}$~yr.  As the tail
extends to 0.48~pc for $d=5$~kpc, this implies $v_f \gapp 6700 (d/5 \;
{\rm kpc})(E / 1 \; {\rm keV})^{1/2} (B_{\rm eq} / 70 \; \mu{\rm
G})^{3/2}$~km~s$^{-1}$.

The tail emission has low flux.  The tail surface brightness in the
2--8~keV band is $4.5 \times
10^{-16}$~erg~s$^{-1}$~cm$^{-2}$~arcsec$^{-2}$, with uncertainty of
$\sim$30\%.  The total unabsorbed flux in the 2--8~keV band in our
extraction region is $9.8 \times 10^{-14}$~erg~s$^{-1}$~cm$^{-2}$, with
similar uncertainty.  The efficiency with which the pulsar's
$\dot{E}$ is converted into tail X-rays in the 2--8~keV band is only
0.00011$(d/5 \; {\rm kpc})^2$, roughly half of the point-source
efficiency.  This is in contrast to other rotation-powered pulsars,
like the Crab, whose X-ray nebular emission is much brighter than the
point-source output.  Without timing information, we cannot rule out an
ultra-compact nebula as the source of the point-source emission.
Several effects  may reduce the X-ray efficiency of ram pressure
confined PWNs.  First,  the efficiency of conversion of  $\dot{E}$ into
X-ray emitting particles may be lower since the reverse shock in the
ram-pressure-confined PWNs is strong only in the forward part of the
head, subtending  a much smaller solid angle than in a static PWN.
Second, a low X-ray efficiency is  expected if the flow time of the
relativistic plasma  through the tail is shorter than  the  synchrotron
life time.  A similar argument was put forth (Chevalier 2000) to
explain the low efficiencies of the Vela and CTB~80 pulsars.  Finally,
low surface brightness emission from beyond the eastern tip of the
observable X-ray tail, or even from G5.27$-$0.90, have gone undetected
in our observation.  Emission from the direction of that nebula, having
X-ray surface brightness half of the ACIS-S3 background, would
contribute roughly two orders of magnitude more flux.

No emission is detected from the shell supernova remnant G5.4$-$1.2.
The upper limits on remnant emission are unconstraining.

\acknowledgments

We thank M. Roberts for discussions.  V.M.K. is a Sloan Fellow
and CRC Chair.  This work is supported by SAO grant GO0-1133A, and by
NASA LTSA grant NAG5-8063 and NSERC grant Rgpin 228738-00
 to V.M.K. E.V.G is supported by the NASA LTSA grant NAG5-22250.
B.M.G. acknowledges a Hubble Fellowship awarded by STScI.


\begin{references}

\reference Becker, W. \& Tr\"{u}mper, J. 1997, \aap, 326, 682
\reference Chevalier, R.~A. 2000, \apjl, 539, L45
\reference Frail, D.~A., Kassim, N.~E., \& Weiler, K.~W. 1994, \aj, 107, 1120
\reference Frail, D.~A. \& Kulkarni, S.~R. 1991, \nat , 352, 785
\reference Gaensler, B.~M. \& Frail, D.~A. 2000, \nat, 406, 158
\reference Kaspi, V. M., Gotthelf, E. V., Gaensler, B. M., Lyutikov, M. 2001, \apjl, 562, L163 
\reference Kennel, C. F., \& Coroniti, F. V. 1984, \apj, 283, 694
\reference Lyne, A.~G. \& Lorimer, D.~R. 1994, \nat, 369, 137
\reference Manchester, R.~N., Kaspi, V.~M., Johnston, S., Lyne, A.~G., \& D'Amico, N.
  1991, \mnras, 253, 7P

\end{references}
\end{document}